\documentclass[14pt]{extarticle}
\usepackage{caption}
\usepackage{hyperref}
\hypersetup{
    colorlinks=true,
    linkcolor=blue, 
    citecolor=malachite
    }
\usepackage{cancel}
\usepackage{ulem}
\usepackage{geometry}
\usepackage[table]{xcolor}
\usepackage[utf8]{inputenc}
\usepackage{mathdots}
\usepackage[T1]{fontenc}
\usepackage[english]{babel}
\usepackage{url}
\usepackage{dsfont}
\usepackage{amsfonts,amsthm,amsmath,amssymb,mathrsfs,bbm,nicefrac}
\usepackage{float} 
\usepackage{cite} 
\usepackage{esvect} 

\definecolor{palatinateblue}{rgb}{0.15, 0.23, 0.89} 
\definecolor{harvestgold}{rgb}{0.85, 0.57, 0.0} 
\definecolor{malachite}{rgb}{0.04, 0.85, 0.32}

\allowdisplaybreaks[1]

\usepackage[colorinlistoftodos]{todonotes}

\usepackage{euscript}
\usepackage{longtable}
\usepackage{graphicx}
\usepackage{color}
\usepackage{shadow}

\usepackage{authblk}

\title{\color{blue}\textbf{\textsc{
Parameterizing Qudit States}}}

\author[1,2,3]{A.~Khvedelidze}
\author[4]{D.~Mladenov}
\author[3]{A.~Torosyan}
\affil[1]{{\normalsize A.Razmadze Mathematical Institute, Iv.Javakhishvili Tbilisi State University, Tbilisi, 
Georgia}}
\affil[2]{{\normalsize
Institute of Quantum Physics and Engineering Technologies, Georgian Technical University, Tbilisi, Georgia}}
\affil[3]{{\normalsize Meshcheryakov Laboratory of Information Technologies, Joint Institute for Nuclear Research, Dubna, Russia}}
\affil[4]{{\normalsize 
Faculty of Physics, Sofia University ``St. Kliment Ohridski'', 
5 James Bourchier Blvd, 1164 Sofia, Bulgaria}}

\date{
}

\begin{document}

\maketitle

\centerline{\textit{In memoriam Vladimir Gerdt {1947-2021}}}
\vspace{0.8cm}

\begin{abstract}\small
Quantum systems with a finite number of states at all times have been a primary element of many physical models in nuclear and elementary particle physics, as well as in condensed matter physics. 
Today, however, due to a practical demand in the area of developing quantum technologies, a whole set of novel tasks for improving our understanding of the structure of finite-dimensional quantum systems has appeared. 
In the present article we will concentrate on one aspect of such studies related to the problem of explicit parameterization of state space of an $N\--$level quantum system. More precisely, we will discuss the problem of a practical description of the unitary $SU(N)\--$invariant counterpart of the $N\--$level state space $\mathfrak{P}_N$\,, i.e., the unitary orbit space $\mathfrak{P}_N/SU(N)$\,. 
It will be demonstrated that the combination of well-known methods of the polynomial invariant theory and convex geometry provides useful parameterization for the elements of $\mathfrak{P}_N/SU(N)$\,.
To illustrate the general situation, a detailed description of $\mathfrak{P}_N/SU(N)$ for low-level systems: 
qubit $(N=2)\,,$ qutrit $(N=3)\,,$ quatrit $(N=4)\,$ \--- will be given. 
\end{abstract}

\newpage

\tableofcontents

\newpage

\section{Introduction}

Quantum mechanics is a unitary invariant probabilistic theory of  finite-dimensional systems. Both basic features, the invariance and the randomness, strongly impose on the mathematical structure associated with the state space $\mathfrak{P}$ of a quantum system. 
In particular, the geometrical concept of the convexity of the state space originates from the physical assumption of an ignorance about  the quantum states. 
Furthermore, the convex structure of the state space, according to the Wigner \cite{Wigner1959} and Kadison \cite{Kadison1965} theorems about quantum symmetry realization, leads to unitary or anti-unitary invariance of the probability measures 
(short exposition of  the interplay between these two theorems see e.g. in \cite{Hunziker1972}).
In turn of the action of unitary/anti-unitary  transformations  
\[\varrho\  \to\  \varrho^\prime = U\varrho U^\dagger \, \]
sets the equivalence  relation $\varrho \simeq \varrho^\prime $ between the states
$\varrho, \varrho^\prime \in \mathfrak{P}$
and defines the factor space $\mathfrak{P}/U$\,.
This space is a fundamental object containing all physically relevant information about a quantum system.
An efficacious way to describe 
$\mathcal{O}[\mathfrak{P}_N]:= \mathfrak{P}_N/ SU(N)\,$ 
for an $N\--$level quantum system is a primary motivation of the present article.  
The properties of $\mathcal{O}[\mathfrak{P}_N]\,,$ as a semi-algebraic variety, are reflected in the structure of the  center of the enveloping algebra $\mathfrak{U}(\mathfrak{su}(N))\,.$ 
Hence, it is pertinent to describe $\mathcal{O}[\mathfrak{P}_N]\,$ using the algebra of  real $SU(N)\--$invariant polynomials defined over the state space $\mathfrak{P}_N\,.$   
Following this observation in a series of our previous publications,
\cite{GKhP2010,GKHP2011,GMPKh2011,GKhP2015,GKhP-X},  we develop description of $\mathcal{O}[\mathfrak{P}_N]\,$  using the classical invariant theory
\cite{PopovVinberg1994}.
On the other hand, $\mathfrak{P}_N/ SU(N)\,$ is related to the co-adjoint orbits space   $\mathfrak{su}^\ast(N)/SU(N)$ and hence it is natural to describe  $\mathfrak{P}_N/
SU(N)\,$ directly in terms of non-polynomial variables \--- the spectrum
of density matrices. 
Below we will describe a scheme which combines these points of view and provides description of the orbit space $\mathfrak{P}_N/SU(N)\,$ 
in terms of one second order polynomial invariant, the Bloch radius of a state and additional non-polynomial invariants, the angles corresponding to the projections of a unit $(N-2)\--$dimensional  vector on the weight vectors of  the fundamental representation of $SU(N)\,.$  

The article is organised as follows. 
The next section is devoted to brief statements of general results about the state space $\mathfrak{P}_N$ of $N\--$dimensional quantum systems, including discussion of its convexity (Section \ref{Sec:convexbody}) and semi-algebraic structure (Section \ref{Sec:semialg}).
Particularly, the set of polynomial inequalities in an $(N^2-1)\--$dimensional Bloch vector and the equivalent set of inequalities in $N-1$ polynomial $SU(N)\--$invariants will be presented for arbitrary $N\--$level quantum systems.
Section \ref{Sec:Orbitspace} contains information on the orbit space
 $\mathcal{O}[\mathfrak{P}_N]$ \---  
the factor space of the state space under equivalence relation against the unitary group adjoint action.
In Section
\ref{ParagtaphQuatrit} we introduce a new type of parameterization of  a qubit,  a qutrit and a quatrit 
based on the representation of  the orbit space of a qudit as a spherical polyhedron on $\mathbb{S}_{N-2}$. 
This parameterization allows us to give a simple formulation of  the  conception of  the hierarchy of subsystems inside one another.
In Section \ref{Sec:NGen} we present formal elements of the suggested scheme for an arbitrary final-dimensional system.
Section \ref{Sec:Conclusion} contains a few remarks on possible applications of the introduced version of the qudit parameterization.

\section{The state space}

The state space of  a quantum  system $\mathfrak{P}_N$ comes in many faces.
One can discuss its mathematical structure from several points of view:  as  a topological set, as  a measurable space, as  a convex body, as  a Riemannian manifold.\footnote{Here is a short and extremely subjective list of publications on these issues \cite{Adelman1993,KusZyczkowski,GrabowskiKusMarmo2005,BengtsonZyczkowski2006}.}
Below we concentrate mainly on a brief description of $\mathfrak{P}_N$ as a convex body realized as a semi-algebraic variety in $\mathbb{R}^{N^2-1}$ following in general the publications \cite{GKhP2010,GKHP2011,GMPKh2011,GKhP2015,GKhP-X}.

\subsection{The state space as a convex body}
\label{Sec:convexbody}

According to the Hilbert space formulation of  the quantum theory, 
a possible state of  a quantum system is associated to a
self-adjoint, positive semi-definite \textit{``density operator''}
acting on  a Hilbert space.
Considering a non-relativistic $N$\--dimensional system whose Hilbert space  
$\mathcal{H}$ is $\mathbb{C}^N\,$, 
the density operator can be identified with the Hermitian, unit trace, positive semi-definite $N\times N$ {\textit{density matrix}~\cite{vonNeumann1927,Landau1927}. 

The set of all possible density matrices forms the \textit{state space} $\mathfrak{P}_N\,$ of an $N$-dimensional quantum system. 
It is a subset of  the space of complex $N\times N$ matrices:
\begin{equation}
    \mathfrak{P}_N=\{ \varrho \in M_N(\mathbb{C})\,|\, \varrho=\varrho^\dagger\,,\ \varrho \geq 0\,,\ \mbox{Tr}\,\varrho= 1 \}\,.
\end{equation}
A generic non-minimal rank matrix $\varrho$ describes the \textit{mixed state}, while the singular matrices with $\mbox{rank}(\varrho )=1$ are associated to \textit{pure states}.
Since the set of $N\--$th order Hermitian matrices has a real dimension $N^2$\,, and due to the finite trace condition $\mbox{Tr}(\varrho)=1$\,,\ 
the dimension of the state space is $\dim(\mathfrak{P}_N)= N^2-1\,.$
The semi-positivity condition $\varrho \geq 0\,$ restricts it further to a certain $(N^2-1)\--$dimensional \textit{convex body}.   
The convexity of 
$\mathfrak{P}_N\,$ is the fundamental property of the state space. 
The next propositions summarise results on  
a general pattern of the state space $\mathfrak{P}_N\,$ as a convex set with an interior $\mbox{Int}(\mathfrak{P}_N)$ and a boundary $\partial\,\mathfrak{P}_N$
\cite{Adelman1993}. 

\noindent{$\bullet\,$\textbf{Proposition I}\,$\bullet\,$} 
\textit{Given two states $\varrho_1\,, \varrho_2 \in \mathrm{Int}(\mathfrak{P}_N)$ 
and a ``probability''  $p\in [0,1]$\,, 
consider the convex combination 
\begin{equation}
    \varrho_p:=(1-p)\varrho_1 + p\varrho_2\,,
\end{equation}
then {$\varrho_p \in \mathrm{Int}(\mathfrak{P}_N)\,.$}}

\noindent{$\bullet\,$\textbf{Proposition II}\,$\bullet\,$} 
\textit{The boundary $\partial \mathfrak{P}_N$ consists of non-invertible matrices of all possible non-maximal ranks: 
\begin{equation}
    \partial\,\mathfrak{P}_N =\{
    \varrho \in \mathfrak{P}_N \,|\, \det(\varrho) = 0\}\,.
\end{equation}
The subset of pure states 
$\mathfrak{F}_N \subset \partial\,\mathfrak{P}_N$\,,
\begin{equation}
\mathfrak{F}_N=\{\varrho \in \partial\mathfrak{P}_N \,|\, \mathrm{rank}(\varrho) = 1\ \}\,,
\end{equation}
contains $N$ extreme boundary points
$\mathscr{P}_i(\varrho)$
which generate the whole 
$\mathfrak{P}_N $ by taking the convex combination:
\begin{equation}
\label{eq:extdecomp}
\varrho = \sum_{i=0}^N\, r_i \mathscr{P}_i(\varrho)\,, \qquad 
\sum_{i=0}^{N}r_i=1\,,\  r_i\geq 0\,.
\end{equation}
In (\ref{eq:extdecomp}) every extreme component $\mathscr{P}_i(\varrho)$ 
can be related to the standard rank-one projector by a common unitary transformation $U\in SU(N)$ and transposition $P_{i(1)}$ interchanging the first and $i$-th position:
\begin{equation}
\mathscr{P}_i(\varrho) = U\,P_{i(1)}\,
\mathbf{diag}(1\,,0\,,\dots\,, 0)\,P_{i(1)}\,U^\dagger \,. 
\end{equation}
}
For any dimension of the quantum system the subset of extreme states provides important information about the properties of all possible states, even the pure states comprise a manifold of a real dimension 
$\dim(\mathfrak{F}_N) = 2N-2\,$, 
smaller than that dimension of the whole state space boundary 
$
\dim(\partial \mathfrak{P}_N)=N^2-2\,.
$  

\subsection{The state space as a semi-algebraic variety}
\label{Sec:semialg}

According to the decomposition (\ref{eq:extdecomp}), the neighbourhood of a generic point of $\mathfrak{P}_N(\mathbb{R}^{N^2-1})\,$  is locally homeomorphic to $\left(U(N)/U(1)^{N}\right)\times D^{N-1}$\,, 
where the component $D^{N-1}$ is an $(N-1)$\--dimensional disc
(cf. \cite{Adelman1993,BengtsonZyczkowski2006}). 
Following this result, below we will describe how the state space $\mathfrak{P}_N$ can be realised as a convex body in $\mathbb{R}^{N^2-1}$ defined via a finite set of polynomial inequalities involving  the Bloch vector of a state. In order to formalize the description of the state space,  we consider the universal enveloping algebra $\mathfrak{U}(\mathfrak{su}(\mathrm{N}))$ of the Lie algebra $\mathfrak{su}(\mathrm{N})\,$.
Choosing the orthonormal basis $\lambda_1, \lambda_2, \dots , \lambda_{N^2-1} $ for  $\mathfrak{su}(\mathrm{N})\,$, 
\begin{equation}\label{eq:su(n)}
 \mathfrak{su}(\mathrm{N})= \sum_{i=1}^{\mathrm{N}^2-1}\,\xi_i\,\lambda_i\,,
\end{equation}
the density matrix will be identified with the element from  $\mathfrak{U}(\mathfrak{su}(\mathrm{N}))$  of the form:
\begin{equation}
\label{eq:denalgebra}
    \varrho(N)= \frac{1}{N}\,\mathbb{I}_{N}+ \sqrt{\frac{N-1}{2N}}\,
     \sum_{i=1}^{\mathrm{N}^2-1}\,\xi_i\,\lambda_i\,.
\end{equation}
The analysis (see e.g. consideration in  \cite{GKhP2010,GMPKh2011}) shows the possibility of description of the state space via polynomial constraints on the Bloch vector of an $N\--$level quantum system. 

\noindent{$\bullet\,$\textbf{Proposition III}\,$\bullet\,$} 
\textit{
If a real $(N^2-1)$-dimensional vector $\boldsymbol{\xi} = (\xi_1, \xi_2, \dots, \xi_{N^2-1}) \,$ in (\ref{eq:denalgebra})
satisfies the following set of polynomial inequalities:
\begin{equation}
\label{eq:ineq}
S_k( \boldsymbol{\xi}) \geq 0\,, \qquad k=1,2, \dots N\,,
\end{equation}
where $S_k(\boldsymbol{\xi})$ 
are coefficients of the characteristic equation of the density matrix $\varrho$: 
\begin{equation}\label{eq:chareq}
\det|| x - \varrho || =  x^N - S_1 x^{N-1} +S_2  x^{N-2}- \dots +(-1)^N\,
S_N  = 0\,, 
\end{equation}
then the equation (\ref{eq:denalgebra})
defines the states $ \varrho\in \mathfrak{P}_N\,$. 
}

The inequalities (\ref{eq:ineq}), 
which guarantee the semi-positivity of the density matrix, remain unaffected by unitary changes of the basis of the Lie algebra and thus the semi-algebraic set (\ref{eq:ineq}) can be equivalently rewritten in terms of the elements of the $SU(N)$-invariant polynomial ring $\mathbb{R}[\mathfrak{P}_N ]^{\mathrm{SU(N)}}\,$. 
This ring can be equivalently represented by the integrity basis in the form of homogeneous polynomials ${\mathcal{P}} = (t_1, t_2, \dots , t_N)\,$, 
\begin{equation}
\mathbb{R}[\xi_1, \xi_2 , \dots, \xi_{N^2-1}]^{\mathrm{SU(N)}}=\mathbb{R}[t_1, t_2, \dots , t_N]\,. 
\end{equation}
The useful, from a computational point of view, polynomial basis ${\mathcal{P}}$ is given by the
\textit{trace invariants}
of the density matrix:
\begin{equation}
\label{eq:traceinvar}
    t_k := \mbox{tr}( \varrho^k )\,.
\end{equation}

The coefficients  $S_k$\,, being $SU(N)$-invariant polynomial functions of the density matrix elements, are expressible in terms of the trace invariants via the well-known determinant formulae:
\begin{equation}
S_k = \frac{1}{k!}\det\left(
\begin{array}{ccccc}
        t_1       & 1              & 0               & \cdots                             & 0    \\
        t_2       & t_1         & 2                & \cdots                             & 1      \\
        t_3       & t_2         & t_1             & \cdots &          \\
        \vdots  & \vdots    & \vdots      & \vdots\, \vdots\,\vdots   & k-1    \\
        t_k       & t_{k-1}   & t_{k-2}    &\cdots                              & t_1
        \end{array}
        \right)\,.
\end{equation}
Aiming at a more economic  description of $\mathfrak{P}_N$\,, we pass from $N^2-1$ Bloch variables to $N-1$ independent trace variables $t_k$. 
The price to pay for such a simplification is the necessity  to take into account additional constraints on $t_k$ which reflect the Hermicity  of the density matrix. 
Below we give the explicit form of these constraints  in terms of
${\mathcal{P}} = (t_1, t_2, \dots , t_N)\,$.  

In accordance with the classical results, the 
\textit{B$\acute{e}$zoutian},
the matrix $\mathrm{B}= \Delta^T\Delta\,,$ constructed
from the Vandermonde matrix $\Delta$\,,
accommodates information on the number of distinct roots (via its rank),  numbers of real roots (via its signature), as well as  the Hermicity condition.  
A real characteristic polynomial has all its roots \textit{real and distinct} if and only if the B$\acute{e}$zoutian is positive definite.  
For generic invertible density matrices \--- matrices with all eigenvalues different, the positivity of the B$\acute{e}$zoutian reduces to the requirement
  \begin{equation}\label{eq:PosBezout}
  \det ||\mathrm{B}||> 0\,.
  \end{equation}
Noting that the entries of the B$\acute{e}$zoutian are simply the trace invariants:
\begin{equation}
\label{eq:Bezoutian}
 \mathrm{B}_{ij}= t_{i+j-2}\,,
\end{equation}
one can be convinced that the determinant of the B$\acute{e}$zoutian is nothing else than the \textit{discriminant} of  the characteristic equation of the density matrix,  
\(
\mbox{Disc} = \prod_{i >j}\left(r_i-r_j\right)^2\,,
\)
rewritten in terms of the trace polynomials
\footnote{
The dependence of the discriminant on trace invariants only up to order $N$ pointed in the left side of
(\ref{eq:discrT}) assumes that all higher trace invariants $t_k$ with $k>N$ in (\ref{eq:discrT})
are expressed via polynomials in $t_1, t_2, \dots,  t_N\,$
(the Cayley-Hamilton Theorem).
}
\begin{equation}\label{eq:discrT}
 \mbox{Disc}(t_1, t_2, \dots, t_{N}): =\det ||\mathrm{B}||\,.
\end{equation}
Hence, we arrive at the following result.

\noindent{$\bullet\,$\textbf{Proposition IV}\,$\bullet\,$} 
\textit{
The following set of inequalities in terms of the trace $SU(N)$\--invariants,
\begin{eqnarray}
\label{eq:Intrace}
 \mathrm{Disc} (t_1, t_2, \dots , t_N ) \geq 0\,, \qquad 
 S_k (t_1, t_2, \dots , t_N )\, \geq 0\,, \qquad 
&& t_1=1\,,
\end{eqnarray}
define the same semi-algebraic variety as the inequalities (\ref{eq:ineq}) in
$N^2-1$ Bloch coordinates do. 
}

\section{Orbit space $\mathfrak{P}_N/SU(N)$}
\label{Sec:Orbitspace}


\subsection{Parameterizing $\mathfrak{P}_N/SU(N)$ via polynomial invariants}

\textbf{Proposition IV} is a useful starting point for establishing a stratification of the  $\mathfrak{P}_N$ under the adjoint action of the $SU(N)$  group. 
It turns out that, due to the unitary invariant character of the  inequalities (\ref{eq:Intrace}), they accommodate all nontrivial information on possible  strata of unitary orbits on the state space $\mathfrak{P}_N$. Indeed, it is easy to find 
the link between the description of $\mathfrak{P}_N$ given in the previous section and  the well-known  method developed by Abud-Sartori-Procesi-Schwarz  (ASPS)  for  construction of the orbit space of compact Lie group  \cite{AbudSartori1983, ProcesiSchwarz1985PHYSLETT,ProcesiSchwarz1985}.
The basic ingredients of this approach can be  very shortly formulated as follows. 

Consider a compact Lie group $G$ acting linearly on a real $d$\--dimensional vector space $V$.
Let $\mathbb{R}[V]^{\mathrm{G}}$ be the corresponding ring of the $\mathrm{G}$\--invariant polynomials on $V$.
Assume ${\mathcal{P}} = \left(t_1, t_2, \dots ,t_q\right)$ is a set of homogeneous polynomials
that form the  integrity basis,
$\mathbb{R}[\xi_1, \xi_2,\dots, \xi_d]^{\mathrm{G}}= \mathbb{R}[t_1, t_2, \dots , t_q]\,.$
Elements of the integrity basis define the polynomial mapping:
  \begin{equation}
  \label{eq:polmap}
t:  \qquad V \rightarrow\mathbb{R}^q\, ; \qquad  (\xi_1, \xi_2,\dots, \xi_d)  \rightarrow (t_1, t_2, \dots , t_q)\,.
\end{equation}
Since the map $t$ is constant on the orbits of $\mathrm{G}$\,, it induces a homeomorphism of the orbit space $V/G$  and the image $X$ of  $t$\--mapping;  $V/G\simeq X$ \cite{CoxLittleO'Shea}.
In order to describe $X$  in terms of ${\mathcal{P}}$  uniquely, it is necessary to take into  account  the
\textit{syzygy ideal} of  ${\mathcal{P}},$  i.e.,
$$
 I_{\mathcal{P}}=\{h \in \mathbb{R}[   y_1, y_2, \dots , y_q]:  h(p_1, p_2, \dots, p_q) =0\,, \  \mathrm{in } \  \mathbb{R}[V ]\, \}.
$$
Let $Z \subseteq  \mathbb{R}^q$ denote the locus of common zeros of all elements of  $I_{\mathcal P}\,,$  then $Z$ is an algebraic subset of $\mathbb{R}^q\,$ such that $X \subseteq Z\,.$ Denoting by  $\mathbb{R}[ Z]$ the restriction of
 $\mathbb{R}[   y_1, y_2, \dots , y_q]$ to $Z$\,, one can easily verify that  $\mathbb{R}[ Z ]$ is
isomorphic to  the quotient\,
\(
\mathbb{R}[   y_1, y_2, \dots , y_q]/I_{\mathcal{P}} \)\, and thus $\mathbb{R}[Z] \simeq \mathbb{R}[V]^{\mathrm{G}}\,.$
Therefore, the subset  $Z$  essentially is determined by $\mathbb{R}[V]^{\mathrm{G}}$,
but to describe  $X $ the further steps are required.
According to  \cite{ProcesiSchwarz1985PHYSLETT,ProcesiSchwarz1985}, the necessary information on $X$ is encoded in the structure of the $q\times q$ matrix  with  elements  given by the  inner products of
gradients, 
$\mathrm{grad}( t_i ):$
\begin{equation}
\label{eq:Grad}
||\mathrm{Grad}||_{ij}= \left(\mathrm{grad}\left(  t_i \right), \mathrm{grad}\left(  t_j \right)\right)\,.
\end{equation}
Hence, applying the ASPS method to the construction of the orbit space $\mathfrak{P}_N/SU(N)$\,, one can prove the following proposition.

\noindent{$\bullet\,$\textbf{Proposition V}\,$\bullet\,$} \textit{
The orbit space  $\mathfrak{P}_N/SU(N)$ can be identified with the semi-algebraic variety, defined as points satisfying two conditions:
\begin{itemize}
\item[a)]
The integrity basis for $SU(N)\--$invariant ring contains only $N$ independent
polynomials, i.e., the syzygy ideal is trivial and the integrity basis
elements of \,$\mathbb{R}[\mathfrak{P}_N]^{\mathrm{SU}(N)}$  are subject to only semi-positivity 
inequalities
\begin{equation}
    S_k (t_1, t_2, \dots , t_N )\, \geq 0\,,
\end{equation}
\item[b)] ASPS inequality  $\mathrm{Grad}(z) \geq 0\,$ is equivalent to the semi-positivity of the  B$\acute{e}$zoutian, provided by existence of the $d\--$tuple where  $\chi=\left( 1,2,\dots, d \right):$
\begin{equation}
\label{eq:Grad2}
\mathrm{Grad}(t_1, t_2, \dots , t_d)= \chi \mathrm{B}\left(
 t_1, t_2, \dots , t_d
\right) \chi^T\,.
\end{equation}
\end{itemize}
}
}
\subsection{
$\mathfrak{P}_N/SU(N)$ \--- as a $\Delta_{N-1}$\--simplex of eigenvalues}

The decomposition of the density matrix  (\ref{eq:extdecomp}) over the extreme states  explicitly displays the equivalence relation between  states, 
\begin{equation}
    \varrho \stackrel{SU(N)}{\simeq} \varrho^\prime \quad \mbox{if} \quad \varrho^\prime = U \varrho\, U^\dagger\,,\quad U \in SU(N)\,.
\end{equation}
Matrices with the same spectrum are unitary   equivalent. Furthermore,  since the eigenvalues of the density matrix $\boldsymbol{r}=(r_1\,,  r_2\,, \dots\,, r_N)$ in (\ref{eq:extdecomp})  can be always  disposed in a decreasing order,  the orbit space $\mathfrak{P}_N/SU(N)$ can be identified with the following ordered  $(N-1)\--$simplex: 
 \begin{equation}
\label{eq:orderedsymplex}
\Delta_{N-1}=\{\boldsymbol{r}\in \mathbb{R}^N\,\biggl|\, \sum_{i=1}^{N}r_i=1\,,
\, 1\geq r_1\geq r_2\geq\dots\geq r_N\geq 0\,\}\,.
\end{equation}

\subsection{
$\mathfrak{P}_N/SU(N)$ \--- as a spherical polyhedron on $\mathbb{S}_{N-2}$}

We are now ready to combine  the above stated methods of the description of the state space $\mathfrak{P}_N$, the polynomial invariant theory and convex geometry to write down a certain parameterization of density matrices.
Based on the extreme decomposition of states (\ref{eq:extdecomp}), the parameterization of the elements of $\mathfrak{P}_N$ is reduced to fixing the coordinates on the flag manifolds of $SU(N)$ and the simplex $\Delta_N$ of eigenvalues of density matrices.  
In the remaining part of the article, we will describe $\mathfrak{P}_N/SU(N)$ in terms of the second order polynomial invariant, which is determined uniquely by the Euclidean length $r$ of the Bloch vector, and $N-2$ angles on the sphere $\mathbb{S}_{N-2}\,,$ whose radius in its turn is given as $\sqrt{\frac{N-1}{N}}\,r$\,.

\subsubsection{Qubit, qutrit and  quatrit}

In order to demonstrate the main idea of the parameterization,  we start with its  exemplification  by considering three the lowest-level  systems, qubit, qutrit and quatrit and afterwards the general case of an $N\--$level system will be briefly outlined.

\paragraph{QUBIT \,$\bullet$}

A two-level system, the qubit, is described by a three-dimensional Bloch vector $\vec{\xi}=\{\xi_1\,, \xi_2\,, \xi_3\}$:
\begin{equation}
\label{eq:qubit}
    \varrho(2) =\frac{1}{2}\,\left(\mathbb{I}_2 +\xi_i\sigma_i\right)\,.
\end{equation}
The qubit state with the spectrum 
$\boldsymbol{r}=\{r_1, r_2\} \in \Delta_1$
is characterized by only one independent second order $SU(2)\--$invariant polynomial $t_2=r_1^2+r_2^2\,.$ 
Introducing the length of the qubit Bloch vector, $r=\sqrt{\xi_1^2+\xi_2^2+\xi_3^2}\,$,
we see that 
\footnote{The semi-positivity of state (\ref{eq:qubit}) dictates the constraint, 
$S_2= 1/2(1 - t_2)\geq 0\,,$ which restricts the value of the Bloch vector length: $0\leq r\leq 1\,.$}
\[
t_2 = \frac{1}{2}+ \frac{1}{2}\,r^2 \,.
\]
Hence, the eigenvalues of the qubit density matrix (\ref{eq:qubit}) can be parameterized as
\begin{equation}
\label{eq:qubitpar}
r_i = \frac{1}{2} +r {\mu}_i\,. 
\end{equation}
It will be explained later that the coincidence of the constants 
$\mu_1= 1/2$ and $\mu_2=-1/2$ in (\ref{eq:qubitpar}) with the standard weights of the fundamental $SU(2)$ representation, when the diagonal Pauli matrix
$\sigma_3$ is used for the  Cartan element of $\mathfrak{su}(2)$ algebra, is not accidental. 
Below we will give a generalization of (\ref{eq:qubitpar}) for the qudit, an arbitrary 
$N\--$level system.
With this aim in mind, it is sapiential to start with considering the $N=3$ and $N=4$ cases. 

\paragraph{QUTRIT\,$\bullet$}

\begin{figure}[H]
\includegraphics[width=23pc]{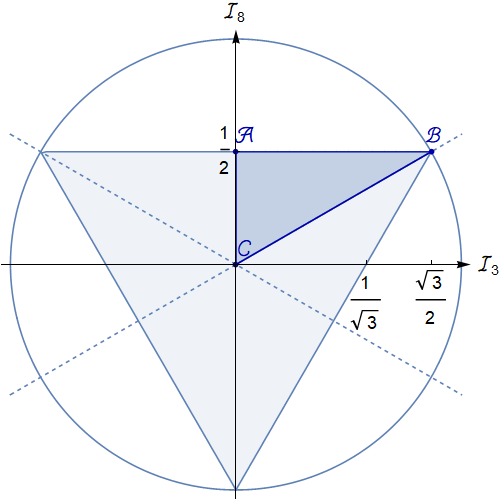}
\hfill
\begin{minipage}[b]{35pc}
\caption{
The $\Delta_2$-simplex of the qutrit eigenvalues is mapped to the triangle $\triangle ABC\,$ 
inscribed in a unit-radius  circle $\mathcal{I}_3^2+\mathcal{I}_8^2=1\,.$
Its inner part $\triangle {ABC}$ comprises the points of the maximal rank-3 states $\mathfrak{P}_{3,3}$
with $1>r_1 > r_2 > r_3> 0\,.$ All these points generate the \textit{regular}  $SU(3)$ orbits  $\mathcal{O}_{123}$ of dimension $\mbox{dim}(\mathcal{O}_{123})=6\,.$ 
The points on the line $AB$ also generate regular orbits $\mathcal{O}_{123}$, however the corresponding states have $\mbox{rank}(\varrho)=2\,.$ 
In contrast to the above case, line 
$AC/\{A\}$ and line $BC/\{B\}$ correspond to the subspace of $\mathfrak{P}_{3,3}\,,$ but now the eigenvalues of the states are degenerate, either  $r_1 = r_2 > r_3\,,$  
or 
$r_1 > r_2 = r_3\,,$ hence representing  the \textit{degenerate orbits} $\mathcal{O}_{1|23}\,$ and $\mathcal{O}_{12|3}$\,, respectively. 
The dimensions of both types of orbits are the same,  $\mbox{dim}(\mathcal{O}_{1|23}) =
\mbox{dim}(\mathcal{O}_{12|3}) =4$\,.
Finally, the single point $C(0,0)$ represents a maximally mixed state which belongs also to the set of rank-3 states.} 
\label{fig:QutritDM-sgn}
\end{minipage}
\end{figure}

We assume that a generic qutrit state $(N=3)$ has the spectrum $\boldsymbol{r}=\{r_1,r_2, r_3\}$ from the simplex $\Delta_2$ and thus is an eight-dimensional object. 
According to the normalization chosen in (\ref{eq:denalgebra}), it is characterized by the 8-dimensional Bloch vector $\vec{\xi}=(\xi_1\,, \xi_2\,, \dots\,, \xi_8)$\,,
\begin{equation}
\label{eq:qutrit}
\varrho(3)=\frac{1}{3}\,\mathbb{I}_3 + \frac{1}{\sqrt{3}}\,\sum_{i=1}^8\, \xi_i\lambda_i\,.
\end{equation}
A qutrit has two independent $SU(3)$ trace invariant polynomials, the first one, $t_2=r_1^2+r_2^2+r_3^2\,$, is expressible via the Euclidean length of the  Bloch vector, $r^2= \sum_{i=1}^8\, \xi_i^2\,,$
\begin{equation}
\label{eq:t2r}
   t_2= \frac{1}{3} + 
   \frac{2}{3} \, r^2\,,
\end{equation}
and the third order polynomial invariant, $t_3=r_1^3 + r_2^3 + r_3^3\,$, which
rewritten in terms of eight components of the Bloch vectors reads:
\begin{eqnarray}
\label{eq:t3qutrit}
t_3= 
    \frac{1}{9} + \frac{2}{3} r^2 
    + \frac{2}{\sqrt{3}} \xi_1 \, (\xi_4 \xi_6 + \xi_5 \xi_7) +\frac{2}{\sqrt{3}}\,\xi_2\, (\xi_5 \xi_6 - \xi_4 \xi_7)
    + \frac{1}{\sqrt{3}}\, \xi_3 \, (\xi_4^2+\xi_5^2-\xi_6^2-\xi_7^2) +
    \frac{1}{9} \xi_8 \, (6 (\xi_1^2+\xi_2^2+\xi_3^2) - 3 (\xi_4^2+\xi_5^2+\xi_6^2+\xi_7^2) - 2 \xi_8^2)\,.
\end{eqnarray} 
Now we want to rewrite 
(\ref{eq:t3qutrit}) in terms of the Bloch vector of a length $r$ and an additional $SU(3)$ invariant. 
Having this in mind, it is convenient to pass to new coordinates linked  to the structure of the Cartan subalgebra of $\mathfrak{su}(3)\,.$ Choosing the latter as the span of the diagonal SU(3) Gell-Mann matrices and noting that 
the state (\ref{eq:qutrit}) is $SU(3)$\--equivalent to the diagonal state:
\begin{equation}
\label{eq:su3I}
    \varrho(3) \stackrel{SU(3)}{\simeq} \frac{1}{3}\,\mathbb{I}_3 + \frac{1}{\sqrt{3}}\,( \mathcal{I}_3\lambda_3 +\mathcal{I}_8\lambda_8)\,,
\end{equation}
one can consider two coordinates $(\mathcal{I}_3\,, \mathcal{I}_8)$ in the Cartan subalgebra of  $\mathfrak{su}(3)$ as independent coordinates in $\mathfrak{P}_3/SU(3)\,.$ 
Taking into account that for the given values of the second trace invariant (\ref{eq:t2r})  the coefficients obey relation  $\mathcal{I}_3^2+\mathcal{I}^2_8=r^2\,, $ we pass to  the polar coordinates on the $(\mathcal{I}_3\,, \mathcal{I}_8)\--$plane, 
\begin{equation}
\label{eq:BlochToPolar}
\mathcal{I}_3= r\, \cos\left(\frac{\varphi}{3}\right)\,, \quad
    \mathcal{I}_8= r\, \sin\left(\frac{\varphi}{3}\right)\,.
\end{equation}
In terms of new variables $(r, \varphi )$ the expression (\ref{eq:t3qutrit}) for the $SU(3)\--$polynomial invariant
$t_3$ simplifies,  
\begin{equation}
\label{eq:t3angle}
t_3= 
\frac{1}{9} + \frac{2}{3}\,r^2 + \frac{2}{9} r^3 \sin{\varphi}\,,
\end{equation}
and the image of the ordered simplex $\Delta_2 $ in the   $(\mathcal{I}_3\,, \mathcal{I}_8)\--$plane under the mapping (\ref{eq:BlochToPolar})   
is given by the triangle $\triangle ABC\,$:
$$
\Delta_2
\mapsto\ \biggl\{\,
 0 \leq \mathcal{I}_3 \leq\frac{\sqrt{3}}{2}\,, \quad 
    \frac{1}{\sqrt{3}}\,\mathcal{I}_3 \leq \mathcal{I}_8 \leq \frac{1}{2}\, \biggl\}\,,
$$
depicted in Figure \ref{fig:QutritDM-sgn}.
The polar form of the invariants (\ref{eq:BlochToPolar}) prompts us to introduce a unit 2-vector \(\vec{n}=(\,\cos(\frac{\varphi}{3})\,, \sin(\frac{\varphi}{3}) \,)
\) and represent the qutrit eigenvalues as 
\begin{equation}
\label{eq:qutritparametr}
r_i = \frac{1}{3} +  
\frac{2}{\sqrt{3}} \, 
r \, \vec{\mu}_i\cdot \Vec{n}\,,
\end{equation}
with the aid of the weights of the fundamental $SU(3)$ representation:
\begin{equation}
 \Vec{\mu}_1=\left(\frac{1}{2}, \frac{1}{2\sqrt{3}}\right)\,,\quad \Vec{\mu}_2=\left(-\frac{1}{2}, \frac{1}{2\sqrt{3}}\right)\,,\quad  \Vec{\mu}_3=\left(0, -\frac{1}{\sqrt{3}}\right)\,. 
\end{equation}
Gathering all together, we convinced that the representation (\ref{eq:qutritparametr}) is nothing else than the well-known  
trigonometric form of the roots of the 3-rd order characteristic equation of  the qutrit density matrix:
\begin{eqnarray}
\label{eq:qutritrig1}
r_1 = \frac{1}{3}-\frac{2}{3}\, r\, \sin\left(\frac{\varphi +4\pi}{3}\right)\,, 
\qquad
r_2 = \frac{1}{3}-\frac{2}{3}\, r\, \sin\left(\frac{\varphi +2\pi}{3}\right)\,, 
\qquad
r_3 = \frac{1}{3} -\frac{2}{3}\, r\, \sin\left(\frac{\varphi}{3}\right)
\,.
\end{eqnarray}

\begin{figure}[H]
\includegraphics[width=.32\textwidth]{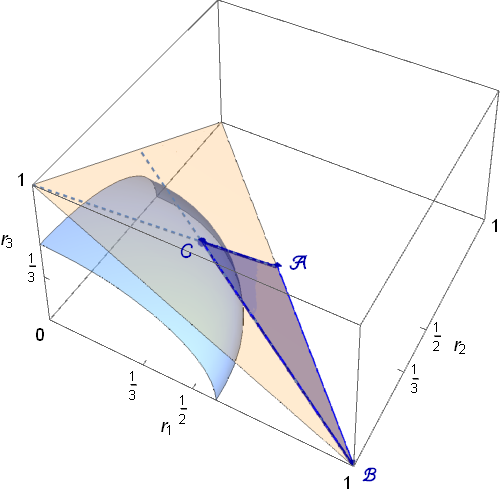}\hfill
\includegraphics[width=.32\textwidth]{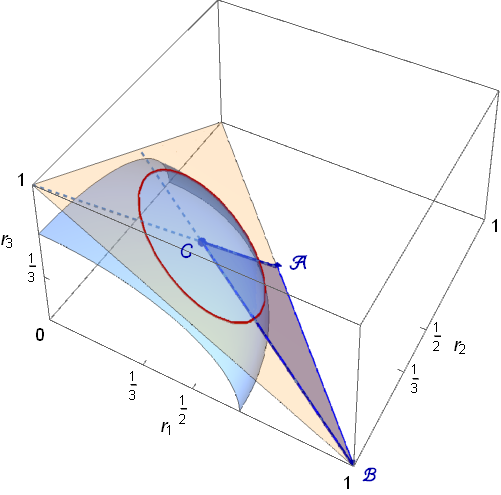}\hfill 
\includegraphics[width=.32\textwidth]{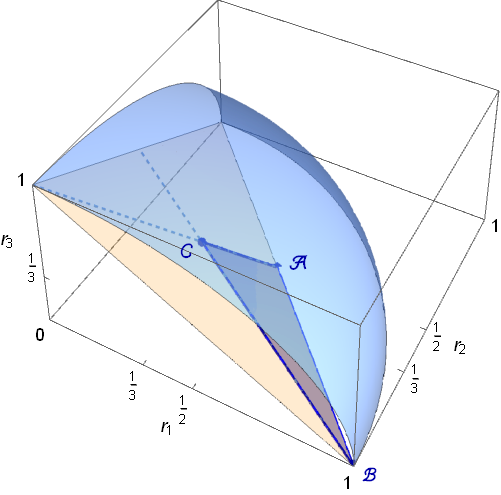}
\caption{\label{fig:ParQutrit}
The picture illustrates a geometrical meaning of the parameterization of qutrit eigenvalues (\ref{eq:qutritrig1}) in terms of the Bloch radius $r$ and the angle $\varphi\in [0,\pi]\,.$ 
Consider an intersection of a qutrit simplex $\Delta_2$ with 2-sphere $r_1^2+r_2^2+r_3^2=\frac{1}{3} + \frac{2}{3}r^2\,.$ 
The intersection depends on the value of a qutrit Bloch vector. For $r=0$ the sphere and the simplex
$\Delta_2$
intersect at point $C=(\frac{1}{3}, \frac{1}{3}, \frac{1}{3})\,,$ while for $0<r< 1$
the intersection is an arc $\mathcal{C}_r$ of a circle on the  plane $r_1+r_2+r_3=1$  of the radius $\sqrt{\frac{2}{3}}\,r\,$ centered at point
$
C(\frac{1}{3}, \frac{1}{3}, \frac{1}{3})\,. 
$
The intersection for $r=1$ takes place at  $B(1,0,0)\,.$
The ordering of eigenvalues $1\geq r_1 \geq r_2\geq r_3 \geq 0$ determines the length of arc $\mathcal{C}_{r}\,.$ 
For any $r$\,, the arc $\mathcal{C}_r$ is  described by  (\ref{eq:qutritrig1}), the  depicted curve in the Figure corresponds to the fixed value $r=1/4$. Furthermore, varying $r$ within the  interval $r\in [0,1]\,,$  provides the 
slices covering the whole simplex $\Delta_2=[0,\pi ]\times \mathcal{C}_r \,.$}
\end{figure}

It is in order to present a 3-dimensional geometric picture associated to the parameterization (\ref{eq:qutritparametr}). 
The three drawings in Figure \ref{fig:ParQutrit} with different values of $r$ show that (\ref{eq:qutritparametr}) are the parametric form of the arc of the red circle  which is the intersection $\Delta_2\cap \mathbb{S}_1(\sqrt{\frac{2}{3}}\,r)\,.$ 

\subparagraph{Qutrit Boundary}

The introduced parameterization is very useful for analyzing the structure of qutrit boundary states. 
The qutrit space $\mathfrak{P}_3$ admits decomposition 
\begin{equation}
\label{eq:qutritSD}
\mathfrak{P}_3= \mathfrak{P}_{3,3}\cup \mathfrak{P}_{3,2}
\cup \mathfrak{P}_{3,1}\,
\end{equation}
into 8d-component of maximal rank-3, 7d-component of  rank-2 and extreme pure states. 
Every component of (\ref{eq:qutritSD}) can be associated with the
corresponding domains in the orbit space $\partial\mathcal{O}[\mathfrak{P}_3]$\,. 
Particularly, the boundary $\partial\mathcal{O}[\mathfrak{P}_3]$ consists of two components and is  described as follows:
\begin{itemize}
\item \textbf{Qubit inside Qutrit} $\bullet$  
For a chosen decreasing order of the qutrit eigenvalues, $r_1\geq r_2\geq r_3,$  the rank-2 states  belong to the edge $\Delta_3\,,$ given by equation $r_3=0\,,$ which in the parameterization (\ref{eq:qutritrig1}) reads: 
\begin{equation}
\label{eq:rank2bound}
\mbox{rank-2~states}:\quad 
\biggl\{
r=\frac{1}{2\, \sin(\varphi/3)}\quad  \text{for} \quad \varphi\in [0,\pi)
  \biggl\}\,.
\end{equation}
Considering (\ref{eq:rank2bound}) as a polar equation for a plane curve, we find that the rank-2 states $\mathfrak{P}_{3,2}$ can be associated to the part of a 3-order plane curve. Indeed, rewriting (\ref{eq:rank2bound}) in Cartesian coordinates $x=r\cos\varphi\,,\, y=r\sin\varphi\,,$ 
\[
(x^2+y^2)(y-3a)+4a^3=0\,,
\]
we identify this curve with the famous Maclaurin trisectrix with a special choice of $a=\frac{1}{2}$.

For the boundary states  (\ref{eq:rank2bound}), the  equations  (\ref{eq:qutritrig1}) reduce to 
\begin{equation}
\label{eq:2qinq3}
r_1= \frac{1}{2}(1+r^\ast_{2\subset3})\,, \qquad
r_2=\frac{1}{2}(1-r^\ast_{2\subset3})\,,
\end{equation}
where 
\begin{equation}
\label{eq:2q3}
r^\ast_{2\subset3} = \frac{2}{\sqrt{3}} \,\sqrt{r^2-
\frac{1}{4}}\,.
\end{equation}
These  expressions for non-vanishing eigenvalues of a qutrit  indicate the  existence of a ``qubit inside qutrit'' whose effective radius is  $r^\ast_{2\subset3}\,.$ 
Since the radius of the Bloch vector of rank-2 states associated to a qubit in qutrit lies in the interval $
 \frac{1}{2}   \leq r < 1\,,
$
the length  of  its Bloch vector, $r^\ast_{2\subset3}$\,, takes the same values as a single isolated qubit, 
$
0\leq r^\ast_{2\subset3} < 1\,.
$   
\item \textbf{Orbit space of pure states of qutrit} $\bullet $ The boundary $\partial\mathcal{O}[\mathfrak{P}_{3,1}]$ corresponding to 
all pure states  $\mathfrak{P}_{3,1}$ is attainable by $SU(3)$ transformation from the point,
$r =1$ for $\varphi=\pi\,.$
\end{itemize}

\paragraph{QUATRIT \,$\bullet $}
\label{ParagtaphQuatrit}

Now, following the qutrit case, consider a 4-level system, the quatrit, whose mixed state is described by the Bloch vector
$\vec{\xi}=\{\xi_1\,, \xi_2\,, \dots,  \xi_{15}\}$, 
\begin{equation}
\varrho(4)=\frac{1}{4}\,\mathbb{I}_4 + 
\frac{3}{2\sqrt{6}}\,\sum_{i=1}^{15}\,\xi_i\lambda_i\,.
\end{equation}
The  integrity basis for a quatrit ring of $SU(4)\--$invariant polynomials
$\mathbb{R}[\xi_1,\dots, \xi_{15}]^{\mathrm{SU}(4)}$
consists of three polynomials $\mathbb{R}[t_2, t_3, t_4]\,.$ Using 
the compact notations (see details in  Appendix \ref{sec:AA}), they can be represented in terms of the Casimir invariants  of $\mathfrak{su}(4)$ algebra in the following form:
\begin{eqnarray}
\label{eq:trinvquatrit}
t_2= \frac{1}{4}+ \frac{3}{4}\,r^2\,,\qquad
t_3= \frac{1}{16}+  \frac{9}{16}\,r^2 +\frac{3}{16}\,\Vec{\xi}\cdot \Vec{\xi}\vee\Vec{\xi}\,, \qquad
t_4=\frac{1}{64}+ 
\frac{9}{32}\,r^2 + \frac{3}{16}\,\Vec{\xi}\cdot \Vec{\xi}\vee\Vec{\xi} +\frac{9}{64}r^4 + \frac{1}{64}\,\Vec{\xi}\vee\Vec{\xi}\cdot \Vec{\xi}\vee\Vec{\xi}\,.
\end{eqnarray} 

\begin{figure}[H]
\begin{minipage}{0.3\textwidth}
    \includegraphics[width=\textwidth]{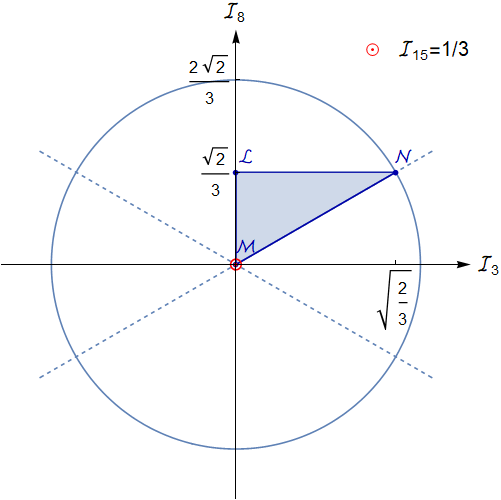}
    \caption{Slice of the convex body (\ref{eq:Ibody}) as a result of  cutting by the plane $\mathcal{I}_{15}=1/3\,.$ }
    \label{Fig:Slice-I3-I8}
\end{minipage}
\hfill
\begin{minipage}{0.68\textwidth} 
From the expressions (\ref{eq:trinvquatrit}) one can see that apart from the length $r$ of the Bloch vector,  there are two independent parameters required to unambiguously characterize the quatrit eigenvalues. To find them, let us proceed as in the qutrit case. Consider the diagonal form corresponding to a  quatrit state:  
\begin{equation}
\label{eq:3diag}
\varrho(4)  \stackrel{SU(4)}{\simeq} \frac{1}{4}\,\mathbb{I}_4 + \frac{3}{2\sqrt{6}}\,( \mathcal{I}_3\lambda_3 +\mathcal{I}_8\lambda_8 +\mathcal{I}_{15}\lambda_{15})\,.
\end{equation}
The coefficients $\mathcal{I}_3\,, \mathcal{I}_8\,$ and $ \mathcal{I}_{15}$ in (\ref{eq:3diag}) are invariants under the adjoint $SU(4)$ transformations of $\varrho$\,.
By equivalence relation (\ref{eq:3diag}),
the quatrit state space is projected to the following convex body:   
 \begin{equation}
\label{eq:Ibody}
 0 \leq \mathcal{I}_3 \leq \sqrt{\frac{2}{3}}\,, 
   \qquad
   \frac{\mathcal{I}_3}{\sqrt{3}} \leq \mathcal{I}_8 
   \leq 
\frac{\sqrt{2}}{3}\,, \qquad
   \frac{\mathcal{I}_8}{\sqrt{2}}\leq \mathcal{I}_{15} \leq \frac{1}{3}\,.  
\end{equation} 
The 2-dimensional slice \,$\mathcal{I}_{15}=1/3\,$ of this body  corresponds to rank-3 states, see Figure \ref{Fig:Slice-I3-I8}.
In terms of new invariants, 
all states with a given length of Bloch vector  $r$ belong to a 2-sphere:
$\mathcal{I}_3^2+\mathcal{I}_8^2 +\mathcal{I}_{15}^2= r^2\,.$
Hence, the corresponding spherical angles $\varphi$  and $\theta$ of these invariants,   
\begin{equation}
\label{eq:Iangles}
\mathcal{I}_3= r \sin\theta \cos\frac{\varphi}{3}\,,
\qquad
\mathcal{I}_8= r \sin\theta \sin\frac{\varphi}{3}\,, \qquad
\mathcal{I}_{15}= r \cos\theta\,,
\end{equation}
can be used as two additional parameters  needed for the parameterization of a quatrit eigenvalues. 
\end{minipage}
\end{figure}

Let us now, in accordance with (\ref{eq:Iangles}), introduce the unit 3-vector   $\Vec{n}= (\sin\theta\cos(\varphi/3)\,,\sin\theta\sin(\varphi/3)\,,
\cos\theta)$  and parameterize 4-tuple of the eigenvalues of the density matrix 
$\boldsymbol{r}=(r_1, r_2, r_3, r_4)$ 
via the following projections:  
\begin{equation}
\label{eq:rhospectparam}
r_i=\frac{1}{4} + 
\sqrt{\frac{3}{2}}
\,r\, \vec{n}\cdot \vec{\mu}_i\,,
\end{equation}
where 3-vectors   $\vec{\mu}_1, \vec{\mu}_2,  \vec{\mu}_3$ and  $\vec{\mu}_4 $ 
denote the weights of the fundamental $SU(4)\,.$
Explicitly the weights read: 
\begin{eqnarray}
    \vec{\mu}_1= \left(\frac{1}{2} 
    \,,  \frac{1}{2\sqrt{3}}
    \,,  \frac{1}{2\sqrt{6}}\right)\,,\qquad
     \vec{\mu}_2=\left(-\frac{1}{2}\,, \frac{1}{2\sqrt{3}}\,, \frac{1}{2\sqrt{6}}\right)\,,\qquad
     \vec{\mu}_3=\left(0, -\frac{1}{\sqrt{3}}\,, \frac{1}{2\sqrt{6}}\right)\,,\qquad
     \vec{\mu}_4=\left(
     0\,, 0\,, -\frac{3}{2\sqrt{6}}\right)\,.
\end{eqnarray}
 Note that the weights $\vec{\mu}_i $ are normalised in a way  leading to a unit norm of the simple roots of algebra  $\mathfrak{su}(4)\,$ and obey relations:
\begin{equation}
\sum_{i=1}^4\vec{\mu}_i=0\,,
\quad
\mbox{and}
\quad
 \sum_{i=1}^4
\,\mu^\alpha_i\,\mu^\beta_i=\frac{1}{2}\,
\delta^{\alpha\beta}\,. 
\end{equation}
Using these expressions, we arrive at the following parameterization of a quatrit eigenvalues:
\begin{align}
\label{eq:4q1}
r_1&=\frac{1}{4} -
\frac{1}{\sqrt{2}}\,r\,\left( \sin\theta \sin\frac{\varphi+4\pi }{3}-\frac{1}{2 \sqrt{2}}\,\cos\theta \right)\,, & r_2&=\frac{1}{4} - 
\frac{1}{\sqrt{2}}\,r\,\left(\sin\theta \sin \frac{\varphi+2\pi }{3}- \frac{1}{2 \sqrt{2}}\,\cos\theta \right)\,,\\
r_3&=\frac{1}{4} - 
\frac{1}{\sqrt{2}}\,r\,\left(\sin\theta \sin{\frac{\varphi}{3}} - \frac{1}{2\sqrt{2}}\,\cos\theta \right) \,, & r_4&=\frac{1}{4}-\, \frac{3}{4}\,r\cos{\theta}\,.
\label{eq:4q4}
\end{align}

\begin{figure}[H]
\includegraphics[width=.32\textwidth]{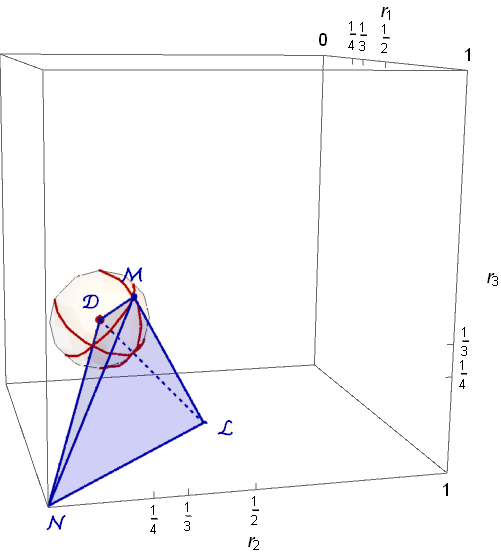}\hfill
\includegraphics[width=.32\textwidth]{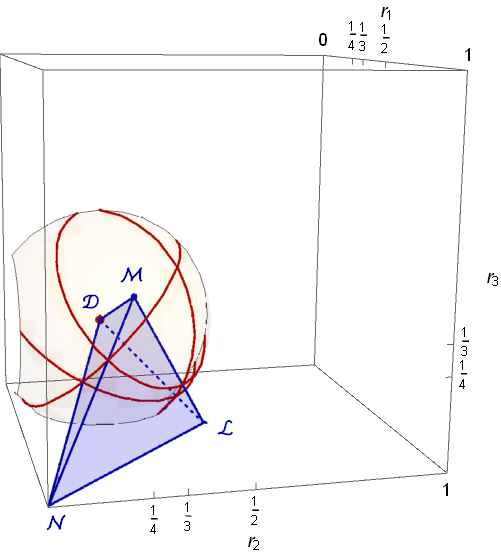}\hfill 
\includegraphics[width=.32\textwidth]{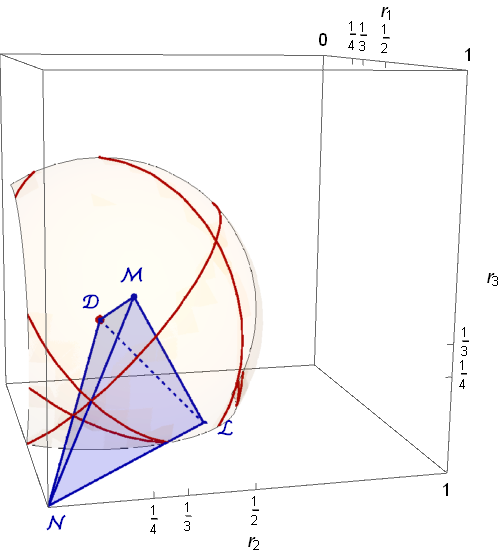}
\caption{A geometric illustration of (\ref{eq:4q1})-(\ref{eq:4q4}).
The  3-sphere $\sum_{i}^4\,r_i^2=1/4+3/4r^2$ intersects the hyperplane $\sum_i^4\,r_i =1\,$ in the positive quadrant.  
The intersection occurs iff\, $\frac{1}{4} \leq 1/4+3/4r^2 \leq 1\,, $  and represents  the  2-sphere $\mathbb{S}_2(\frac{\sqrt{3}}{2} r) $ centered at the point 
$D = (\frac{1}{4}, \frac{1}{4}, \frac{1}{4},\frac{1}{4})\,.$ The intersection with the ordered simplex $\Delta_3$ is given by a spherical polyhedron with 3 or 4 vertices, 
depending on the Bloch radius $r\,.$ 
}
\label{fig:QuatritInter}
\end{figure}

To ensure the chosen ordering of the eigenvalues $r_i \in \Delta_3\,,$   
the Bloch radius should vary  in the interval  $r\in[0,1]\,$ and angles $\varphi, \theta $ be defined over the domains:
\begin{equation}
\label{eq:domaiangle}
 \frac{\pi}{6} < \frac{\varphi}{3} <\frac{\pi}{2}\,,
\qquad 
\cot{\theta} \geq \frac{1}{\sqrt{2}}\sin\left({\frac{\varphi}{3}}\right)\,. 
\end{equation}
A geometric interpretation of  (\ref{eq:4q1})-(\ref{eq:4q4}),
in full analogy with the qutrit case, 
is described  in Figure \ref{fig:QuatritInter}.
The boundary of a quatrit orbit space $\partial\mathcal{O}[\mathfrak{P}_4]$ can be decomposed into 2d-component of  rank-3,  1d-component of rank-2 and extreme zero-dimensional component of rank-1,
corresponding to pure states: \begin{equation}
\partial\mathcal{O}[\mathfrak{P}_4]= \partial\mathcal{O}[\mathfrak{P}_{4,3}]\cup \partial\mathcal{O}[\mathfrak{P}_{4,2}]
\cup\partial\mathcal{O}[
\mathfrak{P}_{4,1}]\,.
\end{equation}

\begin{itemize}
\item \textbf{Qutrit inside Quatrit} $\bullet$ The boundary component $  \mathcal{O}[\mathfrak{P}_{4,3}]$ of rank-3 states
is determined by  the  intersection of 3D simplex $\Delta_3 $ with the hyperplane:
\begin{equation}
\label{eq:r4=0}
r_4=0\,. 
\end{equation}
Parameterizing quatrit eigenvalues in terms of angles, the solution to the equation (\ref{eq:r4=0})  is  
\begin{equation}
\cos\theta =\frac{1}{3\, r}\,, \quad
\text{iff}
\quad r \in [\frac{1}{3}, 1]\,.
\end{equation}
Hence, the parametric form of the 2-dimensional surface 
$  \mathcal{O}[\mathfrak{P}_{4,3}]\,$ is given in terms of the remaining three non-vanishing eigenvalues:
\begin{eqnarray}
\label{eq:3q4q1}
r_1=\frac{1}{3}-\frac{1}{\sqrt{2}}\,f(r)\,
\sin\left(\frac{\varphi+ 4 \pi}{3}\right)\,,\qquad
r_2=\frac{1}{3}-\frac{1}{\sqrt{2}}\,f(r)\,
\sin\left(\frac{\varphi+ 2 \pi}{3}\right)\,,\qquad
r_3=\frac{1}{3}-\frac{1}{\sqrt{2}}\,f(r)
\sin
\left(\frac{\varphi}{3}\right)\,,
\end{eqnarray}
where $f(r)=\sqrt{r^2-\frac{1}{9}}\,.$ 

Consequences of the above derived formulae deserve few comments.
\begin{enumerate}
\item According to the formula  (\ref{eq:3q4q1}) 
for the eigenvalues of boundary rank-3 states, their expressions  are similar to the  
qutrit  eigenvalues given in (\ref{eq:qutritrig1}).  
This observation prompts us to introduce the conception  of the  \textit{``effective  qutrit inside quatrit''}, whose Bloch radius value is determined by the Bloch radius of a quatrit:
\[
r^\ast_{3\subset4} =  \frac{3}{2\sqrt{2}}\,\sqrt{r^2-
\frac{1}{9}}\,.
\]
Note that since the admissible range of the Bloch radius of rank-3 quatrit states is
$
r \in [ \frac{1}{3}, 1]\,,
$
then the effective radius $r^\ast_{3\subset4}$ takes values  in the interval 
$
0 \leq  r^\ast_{3\subset 4} < 1\,.
$


\item  The idea to identify qutrit inside quatrit is based on the establishing correspondence on the level of orbit spaces $\mathfrak{P}_{4,3}$ and $\mathfrak{P}_{3,3}$. 
The generic qutrit state in (\ref{eq:qutritSD}) is 8-dimensional, 
while $\dim(\mathfrak{P}_{4,3})=14\,.$
Thus, one can speak  about the correspondence between quatrit rank-3 states and qutrit states  only modulo unitary transformations.
\item In favour of the idea considering ``effective  qutrit inside quatrit'' is a  relation between the polynomial invariants for states on bulk and boundary.
Particularly, using expressions for trace polynomials 
given in Appendix \ref{sec:AB}., we get: 
\[
t_2^{(4,3)}(r) =
t_2^{(3,3)}(r^{\ast}_{3\subset 4})\,.
\] 
\end{enumerate}

\item 
\textbf{Qubit inside  Qutrit inside Quatrit} $\bullet$
In $\Delta_3 $ the  rank-2 boundary 
component 
$  \mathcal{O}[\mathfrak{P}_{4,2}]$ is comprised from points on a line given by its  intersection  with  two  hypersurfaces:
\begin{equation}
r_4=0\,, \quad r_3=0\,.
\end{equation}
Following in complete analogy with the rank-3 states, we arrive at a ``matryoshka'' structure with 
``effective qubit inside  qutrit which in turn is inside quatrit''. The  Bloch radius of this effective qubit is given by the Bloch radius of a quatrit:
\[
r^\ast_{2\subset 3\subset 4} =  \frac{3}{\sqrt{6}}\,\sqrt{r^2-
\frac{1}{3}}\,.
\]
Note that for rank-2 states $r\in [\frac{1}{\sqrt{3}}, 1]$ and hence $0< r^\ast_{2\subset 3\subset 4}< 1\,.$ 
\item Finally, the rank-1 boundary component $  \mathcal{O}[\mathfrak{P}_{4,1}]$ is generated by one point $\boldsymbol{r}=(1,0,0,0)\, $ which represents all pure states  in $\Delta_3 \,.$ 
\end{itemize}

\subsubsection{Generalization to $N$\--level system}
\label{Sec:NGen}

Now after examining the main features of the introduced parameterization  for a qutrit  and quatrit, we are ready to give a straightforward generalization to the case of an arbitrary $N\--$level system. With this aim, we will use the Cartan subalgebra  of $SU(N)$ as span of the following diagonal $N\times N$ Gell-Mann matrices:
\begin{eqnarray}
\label{eq:SUNCartan1}
&&H_1=
\mathrm{diag}\left(1, -1, 0,\dots,0
\right)\,,\\
\label{eq:SUNCartan2}
&&H_2=\frac{1}{\sqrt{3}}\,
\mathrm{diag}\left(1, 1, - 2,\dots,0
\right)\,,\\
&& \nonumber \ldots\\
\label{eq:SUNCartank}
&&H_k=\frac{2}{\sqrt{2k(k-1)}}\,
\mathrm{diag}\left(\stackrel{k~times}{\overbrace{1,1,\dots, 1}}, -k, 0,\dots,0
\right)\,,\\
\label{eq:SUNCartanN}
&&H_{N-1}=\frac{2}{\sqrt{2N(N-1)}}\,
\mathrm{diag}\left(\stackrel{(N-1)~times}
{\overbrace{1,1,\dots, 1}}, -(N-1)
\right)\,.
\end{eqnarray}
The corresponding weights of the fundamental $SU(N)$ representation  
are
\begin{eqnarray}
    &&\Vec{\mu}_1= \left(\frac{1}{2}\,, \frac{1}{2\sqrt{3}}\,, \dots , \frac{1}{\sqrt{2k(k+1)}}\,, \dots, \frac{1}{\sqrt{2N(N-1)}}\right)\,,\\
    &&\vec{\mu}_2=
    \left(-\frac{1}{2}\,, \frac{1}{2\sqrt{3}}\,, \dots , \frac{1}{\sqrt{2k(k+1)}}\,, \dots, \frac{1}{\sqrt{2N(N-1)}}\right)\,,\\
    &&\vec{\mu}_3=
    \left(0\,, -\frac{2}{2\sqrt{3}}\,, \dots , \frac{1}{\sqrt{2k(k+1)}}\,, \dots, \frac{1}{\sqrt{2N(N-1)}}\right)\,,\\
    && \nonumber \ldots\\ 
    &&\vec{\mu}_k=
    \left(\stackrel{(k-2)~times}
{\overbrace{0,0,\dots,0}}\,, -\sqrt{\frac{k-1}{2k}}\,, \dots , \frac{1}{\sqrt{2k(k+1)}}\,, \dots, \frac{1}{\sqrt{2N(N-1)}}\right)\,,\\
 &&\vec{\mu}_N=
    \left(\stackrel{(N-2)~times}
{\overbrace{0,0,\dots,0}}\,,  \dots , - \sqrt{\frac{N-1}{2N}}\right)\,.  
\end{eqnarray}
It is easy to verify that the following  relations are true:
\begin{equation}
   \sum_{i=1}^N\vec{\mu}_i=0\,,
\quad
\mbox{and}
\quad
 \sum_{i=1}^N
\,\mu^\alpha_i\,\mu^\beta_i=
\frac{1}{2} \, \delta^{\alpha\beta} \,.
\end{equation}
Taking into account these observations, 
one can write down the 
following 
parameterization  for the roots $\boldsymbol{r}$ of the Hermitian $N\times N$ matrix:
\begin{equation}
\label{eq:Nparametr}
    r_i= \frac{1}{N} + 
    \sqrt{\frac{2(N-1)}{N}}\,
    r\,\vec{\mu}_i\cdot \vec{n}\,,
\end{equation}
where $\vec{n} \in \mathbb{S}_{N-2}(1)\,$ and parameter $r$ provides the fulfilment of the correspondence with a value of the second order invariant,  
\begin{equation}
   t_2=\frac{1}{N} + \frac{N-1}{N} \, r^2\,.
\end{equation}
Writing the traceless part of the density matrix as the expansion over the Cartan subalgebra $H$ of $\mathfrak{su}(N)$\,,   
\begin{equation}
\varrho(N) - \frac{1}{N} \,\mathbb{I}_N \stackrel{SU(N)}{\simeq}  \sqrt{\frac{(N-1)}{2N}}\sum_{\lambda\in H }\mathcal{I}_s\lambda_s\,,
\end{equation}
we see that $N-2$ angles of the unit norm vector $\vec{n}$  (\ref{eq:Nparametr})
are related to the invariants $\mathcal{I}^2_3\,, \mathcal{I}^2_8\,, \dots \mathcal{I}^2_{N^2-1}\,,$
whose values are constrained by the Bloch radius $r$\,:
\begin{equation}
\label{eq:moduliWF}
 \sum_{s=2}^{N}\mathcal{I}^2_{s^{2}-1} = r^2\,.  
\end{equation} 
Finally, it is worth to give geometric arguments which emphasise the introduced parameterization (\ref{eq:Nparametr}) of qudit eigenvalues. With this goal
consider the intersection $\mathbb{S}_{N-1}(R) \cap \Sigma_{N-1} $ of $(N-1)\--$sphere of radius $R$ and hyperplane $\Sigma _{N-1}: \sum_i^N r_i=1$ in $\mathbb{R}^{N}\,.$ 
Let us 
describe the hyperplane in  parametric form,   
with parameters $s_1,s_2, \dots, s_{N-1}$\,:
\begin{equation}
\label{eq:plane}
    \boldsymbol{r}= \boldsymbol{d} + 
    \boldsymbol{e}^{(1)} s_1 +  \boldsymbol{e}^{(2)} s_2+ \dots
+ \boldsymbol{e}^{(N-1)} s_{N-1}\,,
\end{equation}
where $N\--$vector $\boldsymbol{d}$  fixes the point $P\in \Sigma_{N-1}$  and the   basis vectors  (Darboux frame) obey conditions: 
$$
\boldsymbol{d}\cdot\boldsymbol{e}^{(\alpha)}=0\,,\quad  \boldsymbol{e}^{(\alpha)} \cdot\boldsymbol{e}^{(\beta)}=\delta^{\alpha\beta}\,, \qquad \alpha, \beta = 1,2, \dots, N-1\,.
$$
Using this parameterization, the equation for $(N-1)\--$sphere is reduced to the constraint 
\begin{equation}
 \boldsymbol{d}^2 + s_1^2+  s_2^2+ \dots
+ s^2_{N-1}= R^2 \,
\end{equation}
for all points of intersection $\mathbb{S}_{N-1}(R) \cap \Sigma_{N-1}\,.$  
Hence, the intersection is nothing else than the $(N-2)\--$sphere of radius $R_{N-2}=\sqrt{R^2-\boldsymbol{d}^2}$ centered at a point associated  to  the vector $\boldsymbol{d}\in \Sigma_{N-1}\,.$
Now if we fix the point $P$ such that $\boldsymbol{d}=(1/N, \dots, 1/N)\,,$ express the  parameters in (\ref{eq:plane}) in  terms of the Bloch radius and the components of the unit vector by relation  $s_\alpha = \sqrt{\frac{2(N-1)}{N}}\, r\, n_\alpha $ and define the  frame vectors $\boldsymbol{e}^{(\alpha)}\,,$ so that
\footnote{Here $\alpha $ component of $i\--$th  weights $\vec{\mu}^{(i)}$ determines    $i\--$th component of basis vector $\boldsymbol{e}^{(\alpha)}\,. $}
\begin{equation}
{e}^{(\alpha)}_i
    = \sqrt{2}\,\mu^{(i)}_\alpha\,, \qquad i=1, 2, \dots, N, \qquad \text{while} \qquad \alpha=1,2,\dots,  N-1\,,
\end{equation}
we arrive at the representation (\ref{eq:Nparametr}) with the radius of intersection sphere  $R_{N-2}=\sqrt{\frac{N-1}{N}}\,r\,.$

Passing from hyperplane $\Sigma_{N-1}$ to its subset, the simplex $\Delta_{N-1}$, we note that $\mathbb{S}_{N-1}(R) \cap \Delta_{N-1} $ will be determined uniquely for every   chosen order of eigenvalues and value of $r\,.$ 
For an arbitrary $N$\,, a special analysis is required to write down 
explicitly $\mathbb{S}_{N-1}(R) \cap \Delta_{N-1}.$ 
Here we only note that the intersection is given by one out of all possible tillings of $\mathbb{S}_{N-2}$ by  the spherical polyhedra. For $N=3$ such polyhedron degenerates to an arc of a circle, whereas for $N=4$  the intersection will be given by 
two types of polyhedra, either  
a spherical triangle, or a spherical quadrilateral, depending on the value of the Bloch radius $r$\,.

\section{Concluding remarks}
\label{Sec:Conclusion}

Since the introduction of the concept of mixed quantum  states, the problem of an efficient parameterization of density matrices in terms of independent variables became one of the important tasks of numerous studies.   Starting with the famous Bloch vector parameterization  
\cite{Bloch1946}, several alternative types of ``coordinates'' for points of quantum states have been suggested \cite{Fano1957,DeenKabirKarl1971,Bloore1976,HioeEberly1981,Dita2005,Akhtarshenas2007,BruningChruscinskiPetruccione,SpenglerHuberHiesmayr,BruningMakelaMessinaPetruccione2012}.
According to the generalization of Bloch vector parameterization, initially introduced for a 2-level system, 
the Bloch vector for an $N\--$level system is a real $(N^2-1)\--$dimensional vector. However, owing to the 
unitary symmetry of an isolated quantum system, those $N^2-1$ parameters can be divided  into two  special subsets. The first subset is given by  $N-1$ unitary invariant 
parameters, and the second one is compiled from the coordinates on a certain  flag manifold constructed from the $SU(N)$ group. 
Introduction  of the coordinates on both subsets has a long history. A description of the former set of $SU(N)\--$invariant parameters is related to the classical problem of determination of roots of a polynomial  equation, while the latter corresponds to a description of the homogeneous spaces of $SU(N)$ group \footnote{Among the important contributions to the problem of parameterizing $ SU (N) $, we would like to mention the following publications that influenced the present work:  \cite{MichelRadicati1973,MacFarline1968,Kusnezov1995,Kortryk2016}.}.

In the present article we have discussed the first part of the problem of parameterization of $N\times N $ density matrices and proposed a general form of parameterization of $N\--$tuple  of its eigenvalues in terms of a  length $r$ of the  Bloch vector and $N-2$ angles on sphere $\mathbb{S}_{N-2}(\sqrt{\frac{N-1}{N}}\,r)$.
We expect that this  parameterization  will be useful from a computational point of view in many physical applications, including models of elementary particles.  
Particularly, in forthcoming publications it will  be used for the evaluation of very  recently introduced indicators of quantumness/classicality of  quantum states  which are  based on the potential of the Wigner quasidistributions to attain  negative values \cite{AKh2017-WF,AKhT-GI2020,AKhT-KZ2021}.

\section{Appendix}
\label{sec:Appendix}

\appendix
\section{Constructing Casimir invariants for $\mathfrak{su}(N)$ algebra }
\label{sec:AA}

In this Appendix we collect few notions and formulae explaining the construction of the polynomial Casimir invariants  on the Lie algebra $\mathfrak{g}=\mathfrak{su}(N)\,$ of the group $G=SU(N)\,.$

Consider algebra $\mathfrak{g} =  \sum_{i}^{N^2-1}\xi_i\lambda_i\,,$ spanned by the orthonormal basis \(\{\lambda_i\}\)
with the multiplication rule 
\begin{equation}
\label{eq:algmult}
    \lambda_i\lambda_j=\frac{2}{N}\,\delta_{ij} +(d_{ijk}+\imath f_{ijk})\,\lambda_k\,,
\end{equation}
defined via the symmetric  $d_{ijk}$ and anti-symmetric  $f_{ijk}$ structure constants. 
Let $\{\omega^i\}$ be the dual basis in $\mathfrak{g}^\ast\,, $ i.e., $\omega^i(\lambda_j)=\delta_j^i$\,, and introduce the $G\--$invariant symmetric tensor $S$ of order $r$:
\begin{equation}
    S=S_{i_1i_2\dots i_r}\,\omega^{i_1}\otimes\omega^{i_2}\dots\otimes \omega^{i_r}\,.
\end{equation}
The $G\--$invariance of tensor $S$ means that 
\begin{equation}
\label{eq:inveq}
  \sum_{s=1}^r \, f^m_{\ ii_s}S_{i_1i_2 \dots i_{s-1} m i_{s+1} \dots i_r}=0\,.
\end{equation}
Using the tensor $S$\,, one can construct 
the elements of the  enveloping algebra 
$\mathcal{U}(\mathfrak{g})\,$:
\begin{equation}
{C}_r= S_{i_1i_2\dots i_r}\lambda_{i_1}\lambda_{i_2}\dots \lambda_{i_r}\,,
\end{equation}
which turns out to belong to the center of 
$\mathcal{U}(\mathfrak{g})\,,$ i.e., 
\(
[{C}_r, \lambda_i] = 0 \)\,,
for all generators\, $\lambda_i\,.$
Having in mind the solution to the invariance equations  (\ref{eq:inveq}), one can build the polynomials in $N^2-1$ real variables $\vec{\xi}=(\xi_1, \xi_2, \dots \xi_{N^2-1})$\,:
\begin{equation}
 \mathfrak{C}_r(\vec{\xi}\,) =\sum_{i}S_{i_1i_2\dots i_r}\,\xi_{i_1}\xi_{i_2}\dots\xi_{i_r}
 \,,
\end{equation}
which are invariant under the adjoint $SU(N)\--$transformations:
\begin{eqnarray}
p(\vv{\mathrm{Ad}_g(\xi)})=p(\vec{\xi}\,)\,. 
\end{eqnarray}
It can be proved that the  symmetric tensors $k^{(r)}$ defined in the given basis of algebra as 
\begin{equation}
 k^{(r)}_{i_1i_2\dots i_r} = \mbox{Tr}\left(\lambda_{\{i_1}\lambda_{i_2}\dots \lambda_{ i_r\}}
 \right)\,, 
\end{equation}
satisfy the invariance equation (\ref{eq:inveq}) and form the basis for the polynomial  ring of $G\--$invariants.
The tensors $k^{(r)}$ admit decomposition  with the aid of the lowest symmetric invariants tensors, $\delta_{ij}$ and $d_{ijk}\,.$ Particularly,  the following combinations are valid candidates
for the basis: 
\begin{equation}
   k^{(4)}_{i_1i_2i_3 i_4}=
   d_{\{i_1i_2 s}
   d_{\{i_3i_4\}s}\,, 
   \qquad k^{(5)}_{i_1i_2i_3 i_4 i_5}=
   d_{\{i_1i_2 s}d_{s i_3 t}
   d_{\{i_4i_5\}t}\,,  
   \qquad k^{(6)}_{i_1i_2i_3 i_4 i_5i_6}=
   d_{\{i_1i_2 s}d_{s i_3 t}
   d_{t,i_4, u}d_{\{i_5i_6\}u}\,.
\end{equation}
As an example,  for $N\--$level system the $G\--$invariant
polynomials up to order six read: 
\begin{eqnarray}
\label{eq:Casimir6}
\mathfrak{C}_2=(N-1)\, {\Vec{\xi}\,}^2\,,\quad
\mathfrak{C}_3=(N-1)\, \Vec{\xi}\cdot\Vec{\xi} \vee\vec{\xi}\,,\quad
\mathfrak{C}_4=(N-1)\, \Vec{\xi}\vee\vec{\xi}\cdot\Vec{\xi} \vee\vec{\xi}\,,\quad 
\mathfrak{C}_5=(N-1)\, \Vec{\xi}\vee\vec{\xi}\vee\Vec{\xi}\vee\vec{\xi} \cdot\Vec{\xi}\,,\quad
\mathfrak{C}_6=(N-1)\, (\,\Vec{\xi}\vee\vec{\xi}\vee\Vec{\xi}\,)^2. 
\end{eqnarray}
In  the equation (\ref{eq:Casimir6}) the  Casimir invariants are represented in a dense vectorial notation using 
the auxiliary $(N^2-1)\--$ dimensional  vector 
defined via the symmetrical structure constants $d_{ijk}$ of the algebra $\mathfrak{su}(N)$\,:
\begin{eqnarray}
(\,\vec{\xi}\vee\vec{\xi}\,)_k := \sqrt{\frac{N(N-1
)}{2}}\,d_{ijk}\,\xi_i\xi_j\,.
\end{eqnarray}

\section{Polynomial $SU(N)\--$invariants on $\mathfrak{P}_N$   }
\label{sec:AB}
In this section the explicit formulae for polynomial invariants for a quatrit will be given in terms of the suggested parameterization of density matrices. 
Since the traceless 
part of the density matrices, 
\(
\varrho-\frac{1}{N}I_N =\sqrt{\frac{(N-1
)}{2N}}\,\, \mathfrak{g}\,, 
\)
belongs to the algebra \(\mathfrak{su}(N)\,\)\,,  
all trace polynomials $t_k$ can be
expanded over the $\mathfrak{su}(N)$ Casimir invariants. The corresponding  decomposition of independent polynomials for the quatrit $(N=4)$ read:
\begin{eqnarray}
t_2=\frac{1}{4}(1+3\mathfrak{C}_2)\,, \qquad
t_3=\frac{1}{4^2}(1+3\mathfrak{C}_2+\mathfrak{C}_3)\,, \qquad
t_4=\frac{1}{4^3}\left( 1+6\mathfrak{C}_2+4\mathfrak{C}_3+
\mathfrak{C}_2^2+\mathfrak{C}_4\right)\,.
\end{eqnarray}
In order to derive the explicit 
form of polynomials $\mathfrak{C}_2$ and $\mathfrak{C}_3$\,, 
the knowledge of components of  the symmetric structure tensor  $d$ is needed. It is convenient at first to express the invariants for diagonal states,  characterized  by $\mathcal{I}_3$\,, $\mathcal{I}_8$ and $\mathcal{I}_{15}\,,$ and afterwards rewrite them for generic states using parameterization (\ref{eq:Iangles}). With this  aim, we collect in Table \ref{table:1}. all non-zero coefficients $d_{ijk}$ (up to permutations) for the  Cartan subalgebra of $\mathfrak{su}(3)$ and $\mathfrak{su}(4)$\,.  
\begin{table}[H]
\centering
\begin{tabular}{ |c|c|c|c|c|c| } 
\hline
 i.j.k & 3.3.8 & 3.3.15& 8.8.8 & 8.8.15 & 15.15.15\\
\hline
\hline
 $d^{\mathrm{SU(4)}}_{ijk}$& $\frac{1}{\sqrt{3}}$& $\frac{1}{\sqrt{6}}$ & $-\frac{1}{\sqrt{3}}$& $\frac{1}{\sqrt{6}}$ &$-\sqrt{\frac{2}{3}}$  \\ 
\hline \vspace{0.02cm}
$d^{\mathrm{SU(3)}}_{ijk}$&$\frac{1}{\sqrt{3}}$&&$-\frac{1}{\sqrt{3}}$&&
\\
\hline
\end{tabular}
\caption{Symmetric structure constants for the Cartan subalgebra of $\mathfrak{su}(3)$ and $\mathfrak{su}(4)$\,.}
\label{table:1}
\end{table}
Taking into account the values for structure constant $d$ from  Table \ref{table:1}.\,, the  Casimir invariants of the third and fourth order of a quatrit  read: 
\begin{eqnarray} 
\label{eq:CasIn43}
&&  \mathfrak{C}_3= 
9 \, \mathcal{I}_{15} \left(\mathcal{I}_3^2+\mathcal{I}_8^2\right) +9 \sqrt{2} \, \mathcal{I}_8 \left(\mathcal{I}_3^2-\frac{1}{3} \, \mathcal{I}_8^2\right) - 6 \, \mathcal{I}_{15}^3 \,,
\\ 
\label{eq:CasIn44}
&&  \mathfrak{C}_4= 
9 \, \left(\mathcal{I}_3^2+\mathcal{I}_8^2\right){}^2  + 36 \sqrt{2} \, \mathcal{I}_8 \,\mathcal{I}_{15} \left(\mathcal{I}_3^2-\frac{1}{3} \, \mathcal{I}_8^2\right) + 12 \, \mathcal{I}_{15}^4\,.
\end{eqnarray}
Finally, plugging   expressions (\ref{eq:Iangles}) into  (\ref{eq:CasIn43})  and (\ref{eq:CasIn44}), we arrive at the  representation of 
the $\mathfrak{su}(4)$ Casimir invariants in terms of quatrit Bloch radius $r$ and 
two angles $(\theta, \varphi)$:
\begin{eqnarray}
\mathfrak{C}_3 &=& 
\frac{3}{4} \, r^3 \left[4 \sqrt{2} \sin ^3(\theta ) \sin (\varphi ) - 3 \cos (\theta ) - 5 \cos (3 \theta )\right]\,,
\\
\mathfrak{C}_4&=& 
\frac{3}{8} r^4 \left[32 \sqrt{2} \sin ^3(\theta ) \cos (\theta ) \sin (\varphi )+4 \cos (2 \theta )+7 \cos (4 \theta )+21\right]\,,
\end{eqnarray}
as well as directly for the trace polynomial invariants,
\begin{eqnarray}
t_2&=& \frac{1}{4} + \frac{3}{4} r^2\,,\\
t_3&=& \frac{1}{16}+ \frac{9}{16} r^2 + \frac{3}{64} r^3 \left(4 \sqrt{2}\sin^3\theta\sin\varphi -3\cos \theta -5\cos(3\theta)\right)\,,\\
t_4&=& \frac{1}{64}+ \frac{9}{32} r^2  + 
\frac{3}{64} r^3 \left(4\sqrt{2}\,\sin ^3\theta \sin\varphi - 3\cos\theta -5\cos (3\theta )\right) +
\frac{3}{512} r^4 \left(32\sqrt{2} \sin ^3\theta \cos \theta \sin\varphi + 4\cos (2 \theta ) +  7\cos (4 \theta ) + 45\right)\,.
\end{eqnarray}


\bigskip
\noindent{\bf Acknowledgements}
The work is supported in part by the Bulgaria-JINR Program of Collaboration. 
One of the authors (AK) acknowledges the financial support of the Shota Rustaveli National Science Foundation of Georgia, Grant FR-19-034. 
DM has been supported in part by the Bulgarian National Science Fund research grant DN 18/3.



\end{document}